\documentclass[journal=jacsat,manuscript=article]{achemso}

\usepackage[version=3]{mhchem} 
\usepackage{textgreek}
\usepackage{gensymb}
\usepackage{subfig}

\title{A synthetic T-cell receptor-like protein behaves as a Janus particle in solution}

\author{Emily Sakamoto-Rablah}
\affiliation[1]
{HH Wills Physics Laboratory, University of Bristol, BS8 1TL, United Kingdom}
\author{Jordan Bye}
\affiliation[2]
{Immunocore Limited, 92 Milton Park, Abingdon OX14 4RY, United Kingdom}
\author{Arghya Modak}
\affiliation[2]
{Immunocore Limited, 92 Milton Park, Abingdon OX14 4RY, United Kingdom}
\author{Andrew Hooker}
\affiliation[2]
{Immunocore Limited, 92 Milton Park, Abingdon OX14 4RY, United Kingdom}
\author{Shahid Uddin}
\affiliation[2]
{Immunocore Limited, 92 Milton Park, Abingdon OX14 4RY, United Kingdom}
\author{Jennifer J McManus}
\email{jennifer.mcmanus@bristol.ac.uk}
\affiliation[1]
{HH Wills Physics Laboratory, University of Bristol, BS8 1TL, United Kingdom}
\alsoaffiliation[3]{Bristol Biodesign Institute, University of Bristol, BS8 1QU, United Kingdom}

\begin{document}

\begin{abstract}
  Protein engineering enables the creation of tailor-made proteins for an array of applications. ImmTACs stand out as promising therapeutics for cancer and other treatments, while also presenting unique challenges for stability, formulation and delivery. We have shown that ImmTACs behave as Janus particles in solution, leading to self-association at low concentrations, even when the averaged protein-protein interactions suggest that the molecule should be stable. The formation of small but stable oligomers has been confirmed by static and dynamic light scattering and analytical ultracentrifugation. Modelling of the structure using Alphafold leads to a rational explanation for this behaviour, consistent with the Janus particle assembly observed for inverse patchy particles.
\end{abstract}

\section{Introduction}
Recent advances in protein engineering have allowed for tailor-made proteins with high selectivity to be developed as therapeutic products \cite{kleinPresentFutureBispecific2024}. This technological advancement holds promise for numerous novel applications, including the design of artificial organelles with custom functions, creating new bio-based materials, or developing new pharmaceutical molecules. Among the latter, one such class of therapeutic molecules is the immune  mobilizing  monoclonal  T-cell receptor against cancer (ImmTAC) \cite{liddyMonoclonalTCRredirectedTumor2012}. ImmTACs target cancerous cells with high specificity by design, while harnessing the body's own immune response and are unique molecules in the biopharmaceutical space. \par
ImmTAC molecules consist of an affinity enhanced T-cell receptor (TCR) joined by a flexible linker to an anti-CD3 single chain variable fragment (scFv) \cite{oatesImmTACs2013, oatesImmTACsTargetedCancer2015}. The TCR region binds to cancer cells with high affinity while the scFv component mobilizes T-cells within the body. This innovative approach holds tremendous potential to deliver life-saving treatments to cancer patients; one ImmTAC molecule (tebentafusp) has FDA approval as a treatment for metastatic uveal melanoma, an aggressive form of eye cancer with previously limited treatment options. As research into ImmTACs (and other synthetic proteins) continues, more options for previously hard-to-treat diseases now seem tractable.\par
The ability to create proteins not found in nature has profound implications for their solution stability. Natural proteins have been finely tuned by evolution to be structurally and colloidally stable under physiological conditions, maintaining a delicate balance between solubility and aggregation to ensure proper cellular function. In contrast, synthetic proteins such as ImmTACs offer a unique opportunity to explore the effects of deliberate design modifications optimised for function, and the consequences for solution stability and therefore formulation and delivery. These modifications, aimed at enhancing target specificity and immune activation, may also impact the protein phase behaviour and self-assembly. \par
Understanding weak, nonspecific protein-protein interactions (PPIs) and self assembly to select formulation conditions is central to realising the transformative potential of these molecules. Ensuring colloidal and structural stability is a crucial aspect of producing safe and effective drug products \cite{wangAntibodyStructureInstability2007, moussaImmunogenicityTherapeuticProtein2016, manningStabilityProteinPharmaceuticals2010}. Colloidal stability is highly dependent on weak, nonspecific PPIs as well as interactions of the protein with solvent and co-solutes. Current manufacturers of biotherapetics often rely on previously successful formulations or use expensive and time-consuming trial-and-error approaches, due to the complex nature of protein intermolecular interactions which has consequences for their phase behaviour. As more novel and innovative biotherapeutics such as ImmTACs are developed for which prior formulation strategies do not exist, established techniques for selecting solution conditions and excipients as routes to formulation need re-evaluation and bringing them to market will require a shift to more rational approaches. \par
Here we demonstrate that the design of a single ImmTAC  (referred to henceforth as ImmTAC1) for biologically optimised function produces a highly anisotropic surface charge distribution, resulting in the unexpected formation of noncovalent oligomers. 

\section{Results and Discussion}
\begin{figure*}[h]
  \includegraphics[width=150mm]{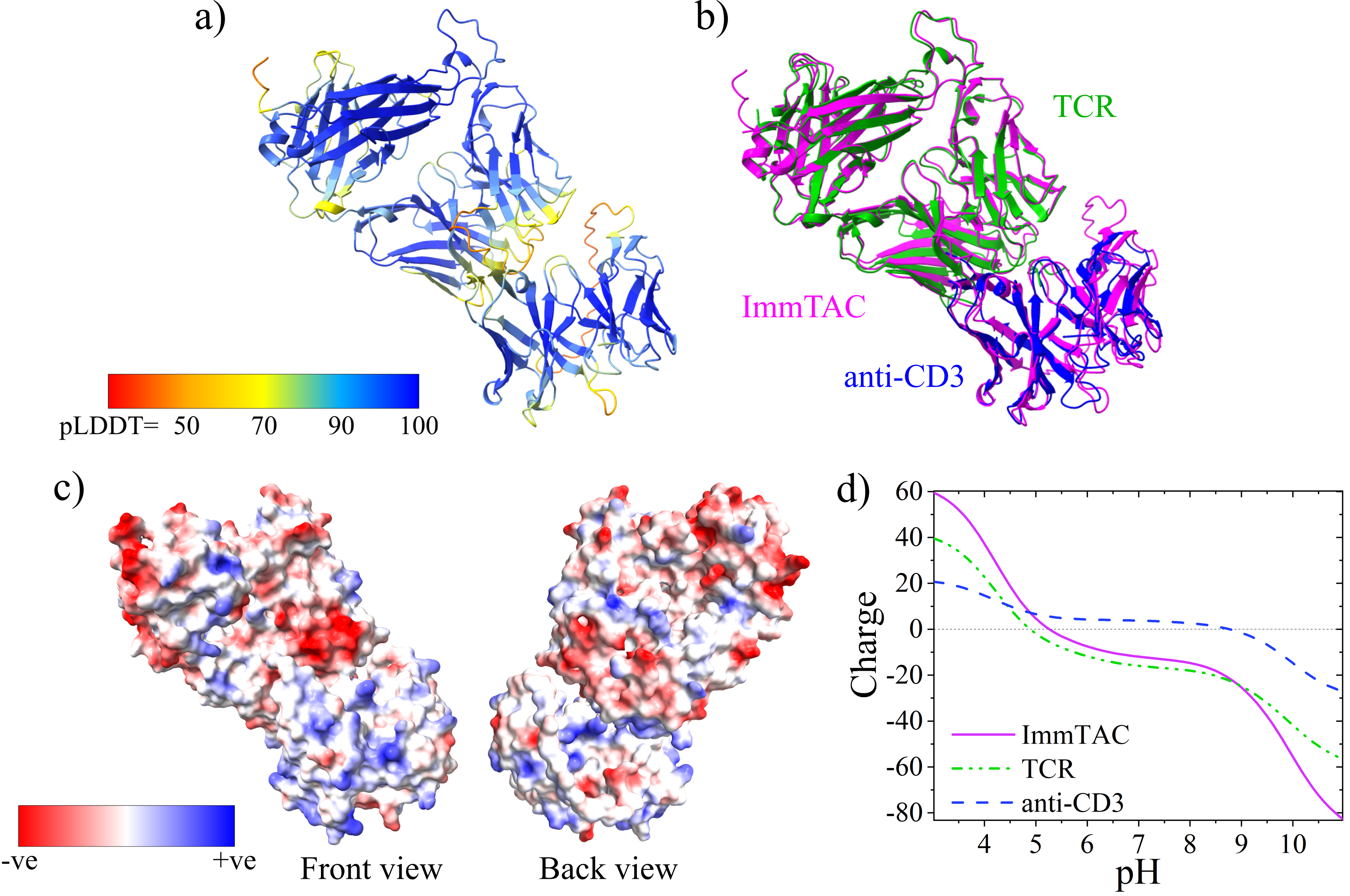}
  \caption{a) Predicted structure of ImmTAC1 generated by Alphafold, coloured by confidence metric pLDDT. b) Alphafold structure (magenta) overlaid with crystal structure of TCR (PDB 2P5E) (green) and anti-CD3 (PDB 1XIW) (blue). c) Front and back view of predicted ImmTAC surface coloured by charge at pH 7.0. Visualisations performed using UCSF ChimeraX \cite{pettersenUCSFChimeraXStructure2021}. d) Estimated total electrostatic charge on ImmTAC, TCR and anti-CD3 regions. Calculated using Prot-pi tool \cite{ProtpiProteinTool}}
  \label{fig:structure}
\end{figure*}
\subsection{ImmTAC structure and structural stability}
Since the stucture of ImmTAC1 has not yet been successfully obtained by crystallography or cryo-EM methods, Alphafold \cite{jumperHighlyAccurateProtein2021} was utilised to generate a predicted structure. Based on the amino acid sequence, five predicted structures were generated. Among these, the structure which ranked highest by predicted local distance difference test (pLDDT) is shown in Figure \ref{fig:structure}(a). The pLDDT is a per-residue confidence metric which estimates how well the prediction would match the real structure based on the local distance difference test \cite{marianiLDDTLocalSuperpositionfree2013}. The highest ranking structure has a total average pLDDT of 86.9\%. As illustrated in Figure \ref{fig:structure}(a), most lower pLDDT regions are those in which we expect structural disorder such as the linkers within the scFv portion, and between the TCR and scFv, rather than areas with significant secondary structure. To further validate the Alphafold prediction, Figure \ref{fig:structure}(b) shows the predicted structure overlaid with known crystal structures for a TCR (PDB: 2P5E) and anti-CD3 fragment (PDB: 1XIW). The predicted structure shows good spatial overlap with the experimental structures, having a root mean squared deviation of 0.624 and 0.672 respectively between the Alphafold structure and PDB structures.\par
The online Adaptive Poisson-Boltzmann Solver (APBS) tool was used to assign electrostatic charges to the Alphafold structure at different pH values \cite{jurrusImprovementsAPBSBiomolecular2018}. A visualisation of the resulting surface charge at pH 7.0 is shown in Figure \ref{fig:structure}(c). The TCR section of the molecule carries a largely negative charge while the anti-CD3 portion is positively charged based on surface exposed charge densities. This is further illustrated by the net charge of the protein at different pH values (calculated using Prot-pi \cite{ProtpiProteinTool}) compared with that of the respective sections (i.e. the scFv and TCR), shown in Figure \ref{fig:structure}(d). This is a feature of the way the protein has been designed and engineered, in that creating a fusion of two proteins with contrasting isoelectric points results in a molecule with a net charge of -12.0 at pH 7, but an anisotropic distribution of that charge.\par
\subsection{Measurement of protein-protein interactions}
\begin{figure*}[h]
    \centering
    \includegraphics[width=130mm]{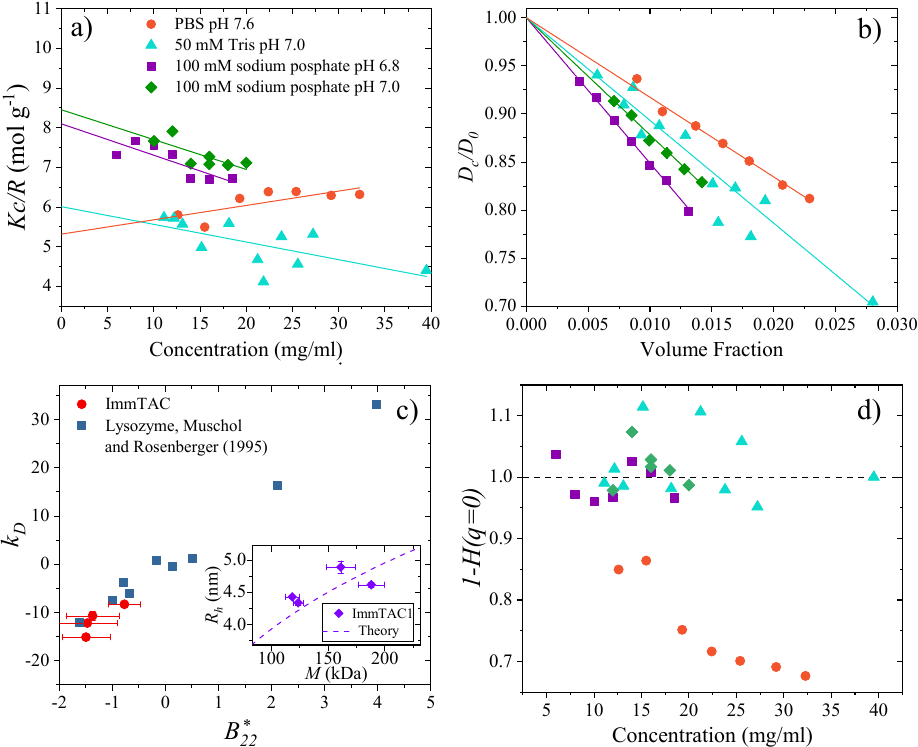}
    \caption{a) SLS data plotted as a Debye plot. Straight lines are linear fits to data with slopes equal to the second virial coefficient $B_{22}$. b) Diffusion data measured by DLS. Straight lines show fits to the data with slope equal to interaction parameter $k_D$. c) Relationship between interaction parameter and second virial coefficient for ImmTAC1 and lysozyme \cite{muscholInteractionsUndersaturatedSupersaturated1995}. Inset shows relationship between measured hydrodynamic radius and molecular weight for ImmTAC1 along with the expected relationship (dashed line) calculated from theory (Equation \ref{equation:Rh_M}). d) Hydrodynamic function calculated using Equation \ref{equation:hydrodynamics} as a function of concentration.}
    \label{fig:LS}
\end{figure*}

To evaluate the strength of net PPIs, static light scattering (SLS) was carried out in 100 mM sodium phosphate at pH 6.8 and 7.0, in phosphate buffered saline (PBS) at pH 7.4, and in 50 mM Tris at pH 7.  The SLS data are shown in Figure \ref{fig:LS}(a) in the form of a Debye plot. We fit a straight line according to the Debye-Zimm equation:
\begin{equation}
    \frac{Kc}{R}=\frac{1}{M} + 2B_{22}c,
\end{equation}
where $c$ is the protein concentration, $R$ is the excess Rayleigh Ratio (defined as the ratio of scattered light to incident light at a given angle, in this case 90\textdegree), $M$ is the protein molecular weight and $K$ is an instrument constant given by 
\begin{equation}
    K = \frac{4\pi^2 n^2 (\frac{dn}{dc})^2}{N_A\lambda^4}
\end{equation}
with $n$ equal to sample refractive index, $\frac{dn}{dc}$ the refractive index increment (here taken to be 0.19 in accordance with Zhao et al \cite{zhaoDistributionProteinRefractive2011}), $N_A$ Avogadro's constant and $\lambda$ the laser light wavelength (632.8 nm). The second virial coefficient $B_{22}$ is in units of volume-moles per mass squared. $B_{22}$ represents deviations to the osmotic pressure introduced by pairwise interactions and hence is often used as a measure of net interaction strength. This is often reported in units of volume, $B_{22}^V =\frac{B_{22} M^2}{N_A} $ and is defined as \cite{muscholInteractionsUndersaturatedSupersaturated1995}
\begin{equation}
    B_{22}^V = -\frac{1}{2}\int_0^\infty[e^{-\frac{w(r)}{k_BT}}-1]4\pi r^2dr,
    \label{equation:B22}
\end{equation}
where $r$ is the protein-protein separation, $w(r)$ is the average interprotein interaction potential, $k_B$ is Boltzmann's constant and $T$ is temperature in Kelvin. According to this definition, positive values of $B_{22}$ correspond to net repulsion between protein molecules (and therefore conditions which are likely conducive to colloidal stability) while negative values reflect net attractive interactions and a likely tendency towards aggregation \cite{muscholInteractionsUndersaturatedSupersaturated1995}. All molecules experience inter-particle repulsion at short range due to steric exclusion, which contributes a positive component to the value of $B_{22}$. Using Equation \ref{equation:B22} and approximating the particles as hard spheres, this excluded volume contribution is $B_{22}^{V,ex} = 4 V_{HS}$ where $V_{HS}$ is the hard sphere volume, i.e. $B_{22}^{V,ex}$ is equal to the excluded volume per particle. Simulations have shown that $B_{22}^{V,ex}$ is well approximated by the excluded volume per particle of a sphere with radius equal to the hydrodynamic radius $R_h$ \cite{grunbergerCoarseGrainedModelingProtein2013}. Here we assume the relation between molecular weight $M$ and radius of gyration $R_g$ to be
\begin{equation}
    \frac{M}{\rho} = \frac{4\pi}{3}R_g^3,
    \label{equation:Rg}
\end{equation}
where $\rho$ is the protein density, taken here to be 1.4 g/cm\textsuperscript{3} \cite{fischerAverageProteinDensity2004}, while we use the molecular weight calculated based on the amino acid sequence of ImmTAC1 to be 77.2 kDa (calculated using the ProtParam tool available on the ExPASy server \cite{gasteigerProteinIdentificationAnalysis2005}). For globular proteins the ratio of $R_g$ and $R_h$ is generally found to be 0.775 \cite{shamirCharacterizationProteinOligomers2019}, hence we estimate
\begin{equation}
    B_{22}^{V,ex} = \frac{4M}{\rho}\left(\frac{1}{0.775^3}\right).
\end{equation}\par
The net contributions of all other pair interactions (i.e. ``soft" interactions) normalised by the hard sphere contributions is often represented as 
\begin{equation}
    B_{22}^* = \frac{B_{22}-B_{22}^{V,ex}}{B_{22}^{V,ex}},
\end{equation}
known as the reduced second virial coefficient. Values obtained from fits to the data in Figure \ref{fig:LS}(a) are shown in Table 1. The fitted $B_{22}^*$ values obtained by SLS indicate net attractive interactions between protein molecules.\par

\begin{table*}[]
\begin{tabular}{|c|c|c|c|c|}
\hline
\textbf{Solution condition} &
  \textbf{$B_{22}^*$} &
  \textbf{\begin{tabular}[c]{@{}c@{}}Molecular\\weight\\ (kDa)\end{tabular}} &
  \textbf{$k_D$} &
  \textbf{\begin{tabular}[c]{@{}c@{}}Hydrodynamic\\radius\\ (nm)\end{tabular}} \\ \hline
PBS pH 7.6                                                               & -0.7 $\pm$ 0.3 & 190 $\pm$ 10 & -8.3 $\pm$ 0.3  & 4.62 $\pm$ 0.04 \\ \hline
50 mM tris pH 7.0                                                          & -1.4 $\pm$ 0.5 & 160 $\pm$ 10 & -10.7 $\pm$ 0.9 & 4.9 $\pm$ 0.1   \\ \hline
\begin{tabular}[c]{@{}c@{}}100 mM sodium\\ phosphate pH 7.0\end{tabular}   & -1.5 $\pm$ 0.6 & 118 $\pm$ 6  & -12.2 $\pm$ 0.4 & 4.42 $\pm$ 0.02 \\ \hline
\begin{tabular}[c]{@{}c@{}}100 mM sodium\\ phosphate pH 6.8\end{tabular} & -1.5 $\pm$ 0.5 & 123 $\pm$ 5  & -15.1 $\pm$ 0.3 & 4.34 $\pm$ 0.03 \\ \hline
\end{tabular}
\caption{Second virial coefficient and molecular weight values measured by SLS and $k_D$ and Hydrodynamic radius values measured by DLS.}
\label{table:LS}
\end{table*}

Dynamic light scattering (DLS) was also used to measure net interaction parameter $k_D$ using the same solution conditions in order to validate the SLS results. The intensity correlation function is given by
\begin{equation}
    G_2(\tau) = \frac{\langle I(t) I(t+\tau) \rangle}{\langle I(t)\rangle ^2},
\end{equation}
where $I$ is intensity of scattered light, $t$ is time and $\tau$ is a lag time, effectively comparing the intensity at a given time $I(t)$ to that at a later time $I(t+\tau)$. For monodisperse solutions the correlation function takes the form \cite{nobbmannDynamicLightScattering2007}
\begin{equation}
    G_2(\tau) = B\left(1 + \beta e^{-2Dq^2\tau}\right) = B\left(1 +  \beta g_2(\tau)\right),
    \label{g2}
\end{equation}
where $D$ is the diffusion coefficient and $q$ is the Bragg wave vector, related to the refractive index $n$ of the solvent, wavelength of scattered light $\lambda$ and the scattering angle $\theta$ by $q=\frac{4\pi n}{\lambda}sin(\theta/2)$. $B$ and $\beta$ are the amplitude (or intercept) and baseline of the correlation function respectively, while $g_2(\tau)$ is referred to as the normalised correlation function. In a polydisperse solution \cite{robertsRoleElectrostaticsProtein2014}, 
\begin{equation}
    g_2(\tau) = \frac{ \sum_{i} \rho_i M_i^2e^{-2D_iq^2\tau}}{ \sum_{i} \rho_i M_i^2}, 
\end{equation}
where $\rho$ is number density and subscript $i$ denotes each species in the solution. We may fit ln($g_2(\tau)$) to a second order polynomial 
\begin{equation}
    ln(g_2(\tau))=-2q^2D_c\tau + q^4 \tau^2(\delta D_c)^2,
\end{equation}
where the first order coefficient gives the collective diffusion coefficient while the second order coefficient corresponds to fluctuations in the diffusion coefficient as described in \cite{robertsRoleElectrostaticsProtein2014}.
From the collective diffusion coefficient we may derive the net interaction parameter $k_D$, using \cite{muscholInteractionsUndersaturatedSupersaturated1995}
\begin{equation}
    \frac{D_c}{D_0} = 1 + k_D \phi,
    \label{equation:kD}
\end{equation}
where $D_0$ is the diffusion coefficient of the protein extrapolated to infinite dilution, and  $\phi = \nu c$ is the protein volume fraction with partial specific volume $\nu$ of the protein taken here to be 0.73 ml/g. Shown in past studies to correlate with the second virial coefficient for some globular proteins \cite{robertsRoleElectrostaticsProtein2014}, a positive $k_D$ value is indicative of net repulsive interactions, while negative $k_D$ indicates net attractive interactions \cite{jamesThermalSolutionStability2012}. However, $k_D$ differs from $B_{22}$ in that there is a contribution from hydrodynamic, or indirect, interactions (since diffusion is affected by both hydrodynamics and thermodynamics) while $B_{22}$ is a purely thermodynamic quantity. Collective diffusion coefficients for ImmTAC1 are shown in Figure \ref{fig:LS}(b). Values of $k_D$ obtained by a linear fit according to equation \ref{equation:kD} are given in Table \ref{table:LS}. In agreement with $B_{22}$ values, the fitted $k_D$ values indicate net attractive PPIs.\par 
In order to test the agreement between the data obtained from SLS and DLS, $k_D$ values were plotted against $B_{22}$, along with values for lysozyme, obtained by Muschol and Rosenberger \cite{muscholInteractionsUndersaturatedSupersaturated1995} and are shown in Figure \ref{fig:LS}(c). Our data show good agreement with the linear relationship measured by Muschol and Rosenberger \cite{muscholInteractionsUndersaturatedSupersaturated1995} within errors (taken to be the error on the gradient given by the linear fits).\par

If we were to only consider the net charge of the protein, this result that interprotein interactions are measured to be net attractive in nature may appear to be somewhat surprising. The pH range studied here (6.8-7.4) is higher than the theoretical isoelectric point of 5.27, with theoretical values for net charge being between -9.8 and -13.5 (Figure \ref{fig:structure}(d)). Based on these values of net charge we might expect double layer repulsion to dominate the net interaction potential at low ionic strengths. Attractive pairwise interaction potentials in charged proteins have been explained in past studies as an attractive electrostatic force between oppositely charged ``patches" on the surface of molecules with anisotropic surface charge density \cite{robertsRoleElectrostaticsProtein2014, yadavInfluenceChargeDistribution2012}. This is consistent with the asymmetry in surface charge observed from the Alphafold models in Figure \ref{fig:structure}(c) and evident in the contrast in charge between the TCR and scFv sections (Figure \ref{fig:structure}(d)). It is reasonable to suggest that this results in electrostatic attraction with the negatively charged TCR and positively charged scFv of two different ImmTACs will experience a charge-charge attraction at sufficiently small intermolecular distances. Indeed, past simulations have also shown the importance of considering anisotropy to explain protein phase behaviour \cite{lomakinAeolotopicInteractionsGlobular1999, altanUsingSchematicModels2019a, robertsRoleElectrostaticsProtein2014, villarSelfAssemblyEvolutionHomomeric2009}.\par
Since DLS measures contributions from hydrodynamic interactions while SLS does not, data from these two techniques may be combined to ascertain the strength of hydrodynamic interactions. The hydrodynamic function $H$ relfects how the drag force on a protein is affected by solvent flow due to surrounding proteins. It can be shown that \cite{muscholInteractionsUndersaturatedSupersaturated1995}
\begin{equation}
    1-H(q=0)=\frac{D_c}{D_0}\frac{R}{KcM}.
    \label{equation:hydrodynamics}
\end{equation}

Resulting values for hydrodynamic functions are shown in Figure \ref{fig:LS}(d). Values for $H$ are seen to be distributed around 0 for all solution conditions other than for PBS at pH 7.6, i.e. hydrodynamic interactions only play a significant role in the protein interactions in PBS. According to the measured $B_{22}$ and $k_D$ values the protein has the least attractive net PPIs in PBS compared with the other buffers examined. PBS contains a high proportion of NaCl, which provides electrostatic screening. This implies that in the case of high screening the direct interactions of electrostatic origin are suppressed sufficiently that hydrodynamic (indirect) interactions have a measurable effect, while under other solution conditions the weak intermolecular electrostatic interactions dominate to the point of suppressing the hydrodynamic interactions.\par
\begin{figure*}[h]
    \centering
    \includegraphics[width=\textwidth]{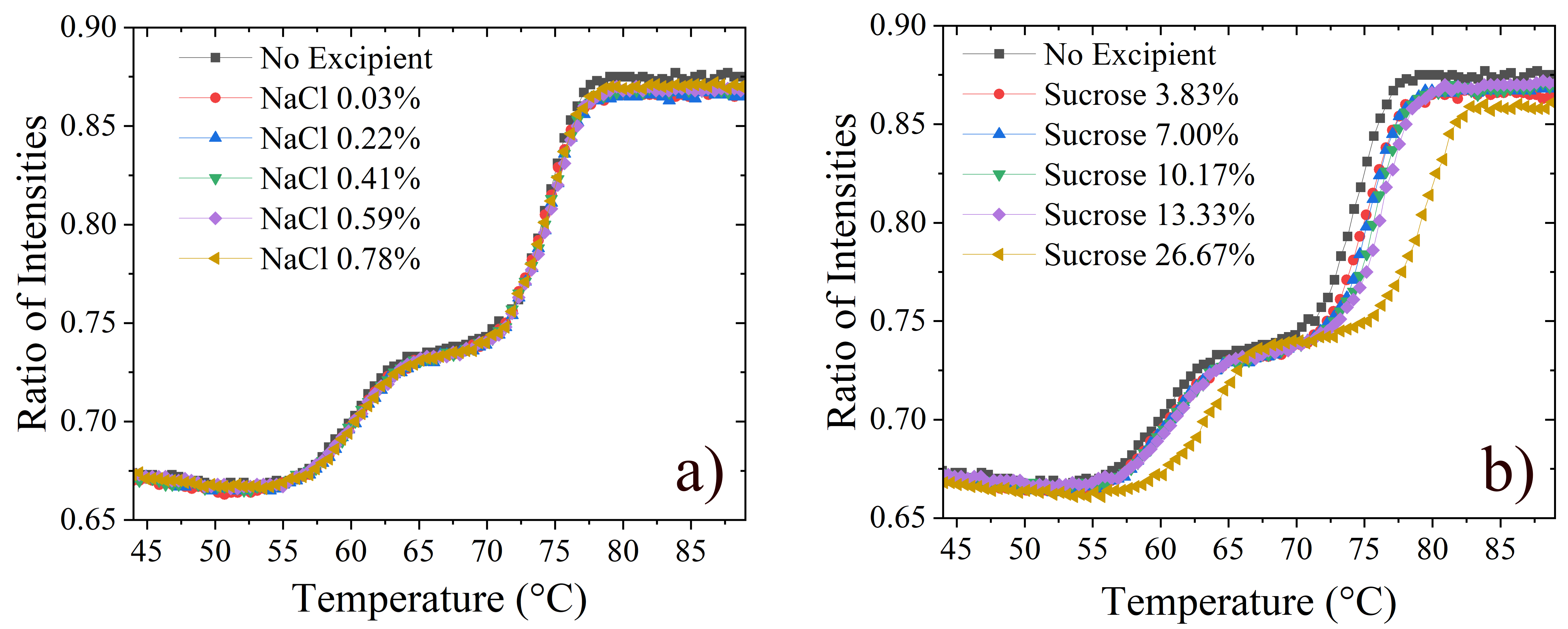}
    \caption{ Thermal unfolding curves measured by nano-DSF in the presence of various concentrations of NaCl (a) and sucrose (b). Concentrations of NaCl and sucrose are in (w/v) percentage.}
    \label{fig:dsf}
\end{figure*}
In order to rule out this difference being due to structural changes such as unfolding, nano differential scanning fluorimetry (nano-DSF) was used to measure structural changes with temperature in the presence of various co-solutes. The temperature of the samples was varied between 20 and 100 \textdegree C and the ratio of fluorescence intensities at 355 and 330 nm was measured in order to monitor the degree of unfolding. The temperature at which ImmTAC1 unfolds ($T_{m}$) was measured in 100 mM sodium phosphate buffer at pH 7.0, shown in Figure \ref{fig:dsf}. The data show two unfolding transitions, corresponding to the scFv section and the TCR section. Data were fitted to the following three-state, biphasic function:
\begin{equation}
     y=
    \frac{(F+m_FT)+K_1(I+m_IT)+K_1K_2(U+m_UT)}{1+K_1+K_1K_2}
\end{equation}
\begin{equation}
    K_1 = e^{-\frac{\Delta H_1}{R}(\frac{1}{T}-\frac{1}{T_{m1}})}
\end{equation}
\begin{equation}
    K_2 = e^{-\frac{\Delta H_2}{R}(\frac{1}{T}-\frac{1}{T_{m2}})}
\end{equation}
where $y$ is the ratio of intensities, $F$, $I$ and $U$ are the intercepts of the lower, intermediate and upper plateaus respectively, $m_F$, $m_I$ and $m_U$ are the gradients of the respective plateaus, $\Delta H_1$ and $\Delta H_2$ are the Van't Hoff enthalpies for the first and second transitions respectively, and $T_{m1}$ and $T_{m2}$ are the corresponding unfolding temperatures, defined as the temperature values halfway between states. In buffer, the transition temperatures produced by the fit are 59.80 $\pm$ 0.05 \textdegree C and 74.28 $\pm$ 0.02 \textdegree C, corresponding to the unfolding of the two sections. The $T_m$ values indicate a high degree of structural stability, likely due to the addition of a non-native disulphide bond in the design of the ImmTAC molecules \cite{boulterStableSolubleHighaffinity2005}. The addition of sucrose had the effect of increasing the thermal stability, with a continuous increase in $T_m$ values as sucrose concentration was increased; the highest concentration measured here (26.6 \% (w/v) sucrose) increased the unfolding temperatures by 3.48 \textdegree C and 4.82 \textdegree C respectively (Figure \ref{fig:dsf}(b)). The presence of NaCl (Figure \ref{fig:dsf}(a)) had no effect on the structural stability of the protein indicating that differences in $\Delta H$ between conditions with and without NaCl is not due to changes in structure. The highest NaCl concentration measured was 0.78 \% (w/v), which is comparable to the NaCl concentration in PBS buffer. 

\subsection{Observation of oligomeric states}

\begin{figure*}[h!]
 \centering
 \includegraphics[width=\textwidth]{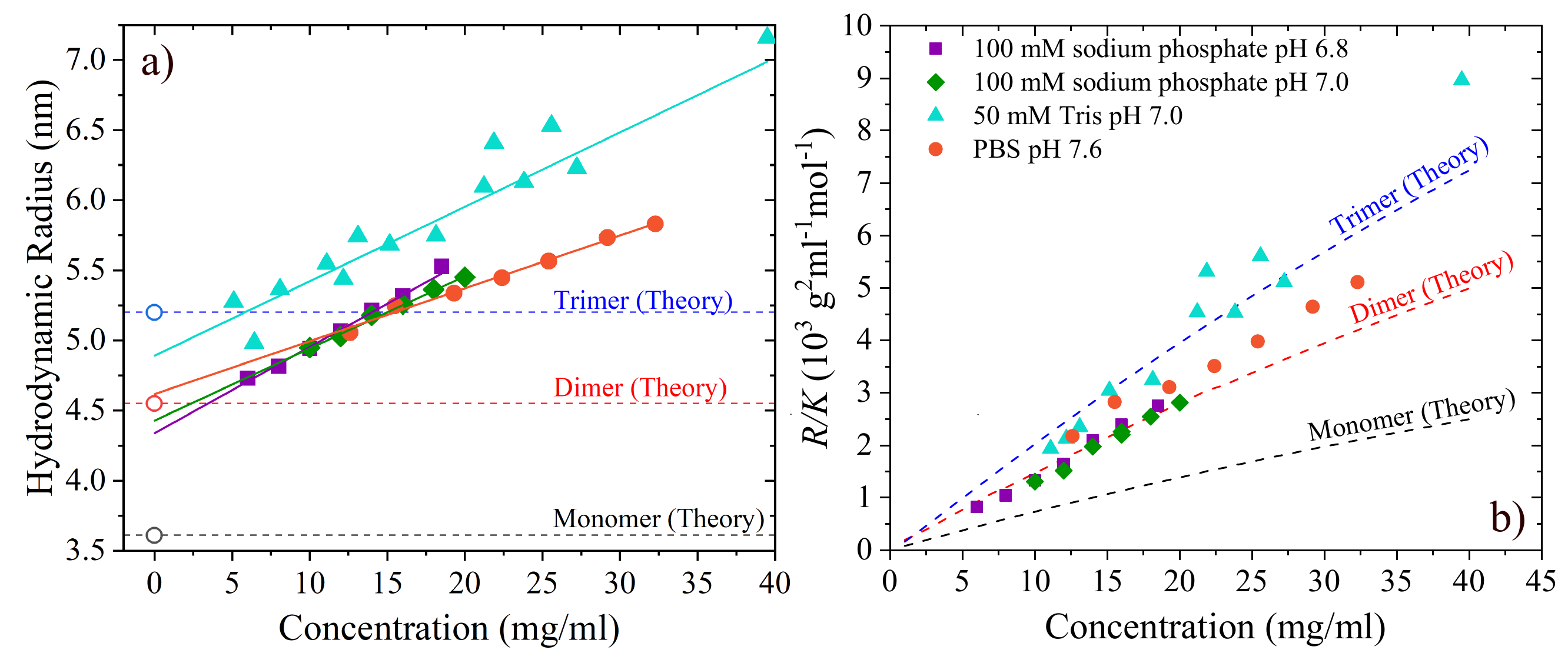}
 \caption{a) Hydrodynamic radius of ImmTAC1 measured by DLS as a function of concentration. Solid lines are linear fits to the data. Open symbols represent predicted hydrodynamic radius values at infinite dilution (calculated using Equation \ref{equation:Rh_M}) for monomer, dimer and trimer respectively. b) Scattered light intensity as a function of concentration. Dashed lines are theoretical scattering intensities for monomer, dimer and trimer respectively.}
 \label{fig:size}
\end{figure*}
Using diffusion data obtained using DLS, hydrodynamic radius may be calculated using the Stokes-Einstein equation
\begin{equation}
    D = \frac{k_B T}{6\pi \eta R_h}
\end{equation}
where $\eta$ is the viscosity of the solvent. Resulting radii are plotted in Figure \ref{fig:size}(a) with a linear fit. Using Equation \ref{equation:Rg} and again assuming that $R_g/R_h=0.775$, we estimate the theoretcial hydrodynamic radius to be
\begin{equation}
    R_h = \frac{1}{0.775}\left(\frac{3}{4\pi}\frac{M}{\rho}\right)^{1/3}.
    \label{equation:Rh_M}
\end{equation}
This relationship is in good agreement with experimental data reported by Smilgies and Folta-Stogniew \cite{smilgiesMolecularWeightGyration2015}, who compared protein molecular weights and hydrodynamic radius values obtained by DLS. Theoretical hydrodynamic radii for an ImmTAC1 monomer, dimer and trimer are shown as dotted lines in Figure \ref{fig:size}(a). Measured values of $R_h$ extrapolated to infinite dilution lie between theoretical values for monomer and trimer, suggesting that there are some oligomers present even at the lowest concentrations. In Figure \ref{fig:size}(b), scattered light intensities measured by SLS are plotted along with theoretical scattering intensities for monomer, dimer and trimer solutions respectively, calculated from scattering theory as detailed by Minton \cite{mintonStaticLightScattering2007}. In agreement with the hydrodynmamic radii, the measured scattering intensities suggest that the average particle size is larger than monomer but (mostly) smaller than trimer. It is important to note that both of these techniques give an estimation of ``average" size, for example a measurement of $R_h$ or scattered intensity corresponding to ``trimer" could in fact result from a mixture of monomer and a range of small oligomers, for example. \par
\begin{figure*}[h]
 \centering
 \includegraphics[width=120mm]{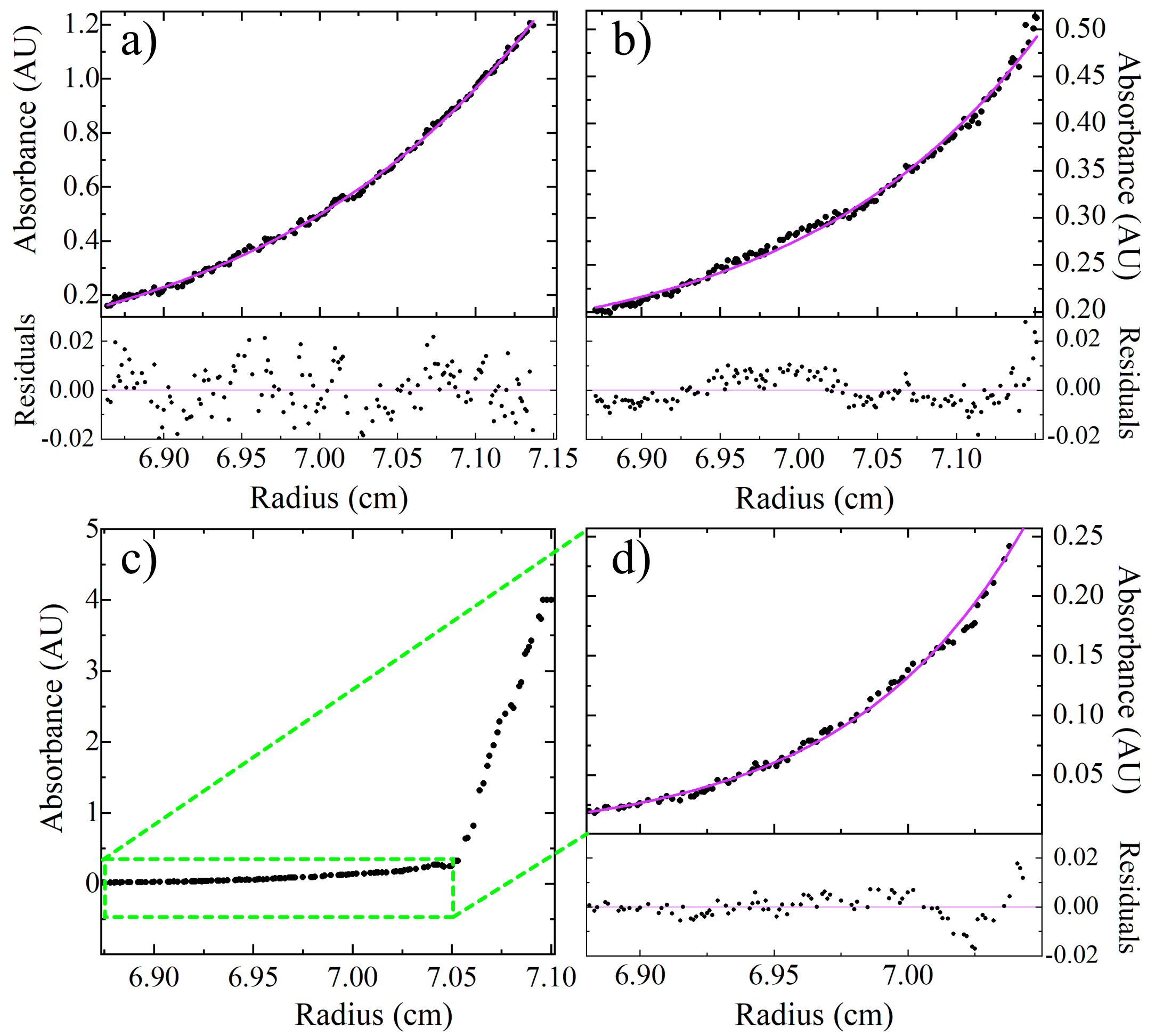}
 \caption{Sedimentation equilibrium AUC data for ImmTAC1 measured a) 1 mg/ml at 9000 RPM b) 11 mg/ml at 6000 RPM and c),d) 11 mg/ml at 15,000 RPM. Purple lines are fits to the data. }
 \label{fig:AUC}
\end{figure*}

To determine the oligomer sizes consistent with the DLS and SLS data, we employed sedimentation equilibrium analytical ultracentrifugation (SE-AUC). Data were fitted using the SEDPHAT software \cite{schuckAnalysisProteinSelfassociation2003a} to the equation for a single ideal non-interacting species \cite{coleAnalyticalUltracentrifugationSedimentation2008}
\begin{equation}
    c(r) = c_0 exp \left[ \frac{M_b\omega^2}{RT}\left(\frac{r^2-r_0^2}{2}\right) \right],
\end{equation}
where $c_0$ is the concentration at a reference radius $r_0$, $\omega$ is the rotor speed and $R$ is the gas constant. $M_b=M_w(1-\nu\rho_s)$ is the buoyant mass of the particle, with $\nu$ the partial specific volume of the particle and $\rho_s$ the solvent density. At 1 mg/ml (Figure \ref{fig:AUC}(a)), the data show a good fit to a single ideal species with a fitted molecular weight of 78.00 kDa, in good agreement with the the theoretical molecular weight for a monomer based on the amino acid sequence (77.2 kDa). At 11 mg/ml (Figure \ref{fig:AUC}(b)), attempting to fit the data to a single ideal species results in a non-ideal distribution of residuals, pointing to some polydispersity. The molecular weight from the fit was 209.70 kDa (2.72 times the theoretical monomer weight). When the 11 mg/ml sample was centrifuged at a higher speed of 15,000 RPM, the data showed two distinct sedimentation pools (Figure \ref{fig:AUC}(c)). Analysing the the lower molecular weight species (Figure \ref{fig:AUC}(d)), there is a good fit to a single species with a fitted molecular weight of 80.10 kDa. This suggests that a non-negligible amount of monomer is present, along with a higher-order oligomer, with the total molecular weight pointing towards trimer.\par
Given that our data indicate that net PPIs are dominated by electrostatic attraction, the formation of small oligomers is a reasonable conclusion. We may rule out partial unfolding or other structural changes as the cause of oligomerization, since the unfolding temperatures and unfolding enthalpies that we measured (Figure \ref{fig:dsf} and Supplementary information)  indicate a high degree of structural stability at room temperature, with the stability not significantly enhanced by the addition of common stabilizing excipients. It is unlikely that there are any structural changes occuring at the temperature (20 \textdegree C) at which the DLS and SLS were carried out. This further supports that the assembly into higher order structures is due to the PPIs rather than any partial unfolding of the protein structure. \par

While reports of anisotropic charge distributions leading to net-attractive PPIs for proteins are rare, more work has been explored for colloidal systems where ``patchy" particles have been studied widely by coarse-grained simulations \cite{kressColloidalInteractionsGet2020, vissersPredictingCrystalsJanus2013, sciortinoPhaseDiagramJanus2009}. Janus particles are particles with two distinct surfaces or regions with different properties. Charged Janus particles, colloids exhibiting positive charge on one side and negative charge on the other have come to be known as inverse patchy particles (IPCs). This may be thought of as the simplest possible coarse-grained model of the surface charge distribution of ImmTAC1, which has a positively charged end and a negatively charged end (upper section of the molecule in Figure \ref{fig:structure}(c)). Simulations of IPCs have been shown produce multiple phases including crystals, gels and clusters \cite{dempsterAggregationHeterogeneouslyCharged2016, hongClustersChargedJanus2006b, kStabilizingOrderedStructures2021} ranging from trimer to higher order multimers. At lower concentrations, size-limited clusters are formed, which this supports the self assembly observed here.

The question remains as to why the protein appears to form predominantly trimers (by AUC) and, on the timescales of our experiments, do not proceed to form higher order oligomers. In all conditions measured, the polydispersity index (PDI) is smaller than 0.18, and in all conditions other than Tris, is smaller than 0.07 (see Supplementary Information), indicating that no large aggregates are present. A species with at least double the hydrodynamic radius generally needs to be present in order to see the polydispersity reflected in an increased PDI value, which corresponds to an eight-fold increase in molecular weight. This indicates that the oligomers being formed are smaller than 8-mer. One possible explanation is the asymmetry between the sizes of the two charged regions. The TCR section which carries a charge of -15.9 at pH 7.0 (see Figure \ref{fig:structure} (d)) has a solvent accessible surface area (SASA) of 215.2 nm\textsuperscript{2} while the anti-CD3 section carries a charge of 3.8 at the same pH with a SASA of 121.2 nm\textsuperscript{2} (SASA values were calculated using the ``measure sasa" tool available in UCSF ChimeraX \cite{goddardUCSFChimeraXMeeting2018}). It has been shown in simulations of both spherical and circular IPCs that as the difference in size between the two regions increases, the number of attractive contacts that each particle can make without forming repulsive contacts decreases, leading to lower coordination number and limiting cluster sizes \cite{dempsterAggregationHeterogeneouslyCharged2016, kStabilizingOrderedStructures2021}. Another possible explanation is that once a small oligomer is formed, the total charge is sufficient to stabilize the oligomer. Studies have reported that increasing the charge of a protein, either through mutanegnesis \cite{austerberryEffectChargeMutations2017, lawrenceSuperchargingProteinsCan2007, simeonovSurfaceSuperchargedHuman2011} or nonspecific binding of small charged molecules to the surface \cite{byeATPTriPolyphosphateTPP2021, zalarNonspecificBindingAdenosine2023} resutls in an increase of the colloidal stability of the molecule in solution by increasing the intermolecular electrostatic repulsion at low ionic strength. Hence, we hypothesize that upon small oligomer formation, the local charge anisotropy is reduced, producing an oligomer with a more balanced distribution of charge across the surface and a more isotropic net interaction potential. \par
It is interesting to note that during preparation of all samples (for both light scattering and AUC experiments), the highest concentration samples were prepared first and diluted to obtain the lower concentrations. During AUC, the oligomers observed in the 11 mg/ml sample were not observed in the 1 mg/ml sample, indicating that the higher order structures had dissociated in this diluted sample. However, the hydrodynamic radii measured by DLS close to 1 mg/ml indicate the presence of species larger than monomer, implying that dilution alone is not enough to cause the oligomers to dissociate. Indeed, the fact that centrifugation of the 11 mg/ml sample at higher speeds was able to remove the oligomeric species altogether leaving monomer behind suggests that there is no association/dissociation equilibrium between monomers and small oligomers, at least on shorter timescales. It is possible that the formation of trimer is weakly reversible, and that shear forces from the ultracentrifugation itself is enough to cause dissociation at low concentration.

\section{Conclusions}
We have shown that ImmTAC1 exhibits an anisotropic distribution of surface charge akin to a Janus particle, due to its design as a fusion of two proteins with different isoelectric points. Due to this feature, the molecule exhibits net attractive interactions, dominated by electrostatic interactions between regions of unlike charges on different molecules. This in turn leads to the self assembly of small oligomers, which coexist with monomers.\par
This work has implications for the rational design of protein and/or peptide self-assembly; it may be feasible to direct desired self-assembled states by engineering molecules with different configurations of anisotropic surface charge.\par
Understanding the impact of protein structure and anisotropy on PPIs and self assembly is also important for formulation; ultimately these proteins are designed to be therapeutic molecules and as such require stable solution conditions in order to be used for treatment. Understanding the solution behaviour of anisotropic molecules as Janus particles opens the door for more rational approaches to formulation and selection of formulation excipients in the future.

\section{Methods and Materials}
\subsection{Sample Preparation}
ImmTAC molecules were provided by Immunocore as 2.4 mg/ml solutions in PBS. Reagents were purchased from Thermo Fisher Scientific and Merck-Millipore. Buffers containing 100 mM sodium phosphate at pH 6.8 and 7, 50 mM Tris at pH 7.0 and PBS at pH 7.6 were all prepared in Milli-Q water (system by Merck Millipore) by dissolving the appropriate amount of salts. All buffers contained 0.02\% w/v sodium azide to prevent microbial growth and were degassed and filtered using a 0.45 $\mu$m, 47 mm cellulose filter (Millex or Sartorius). Amicon Ultra-4 (10 kDa molecular weight cutoff) from Merck-Millipore were used to concentrate protein solutions and buffer exchange to the appropriate buffer. Protein concentrations were measured by UV spectroscopy at a wavlength of 280 nm.

\subsection{Light scattering}
Static and dynamic light scattering (SLS and DLS) were performed using an ALV/CGS-3 goniometer and ALV/LSE-5004 Multiple Tau Digital correlator with a 638.2 nm HeNe laser. The sample environment was kept at 20 \textdegree C using a Thermo Scientific DC30-K20 water bath, and measurements were taken at a 90\textdegree scattering angle. Protein samples were centrifuged at 13,000 xg for one hour prior to measurements in order to remove aggregates. Measurements were made starting with the highest concentration and performing a serial dilution to obtain the lower concentrations.
\subsection{Analytical Ultracentrifugation}
Sedimentation equilibrium analytical ultracentrifugation (SE-AUC) was performed using a Beckman XL-I analytical ultracentrifuge with dual sector epon-filled centrepieces with quartz glass in an AN-50 Ti rotor. Samples were sedimented at 1907, 4292 and 11924 xg at concentrations of 1 and 11 mg/ml.
\subsection{Nano Differential Scanning Fluorimetry}
 Nano differential scanning fluorimetry (Nano-DSF) was carried out using an Applied Photophysics SUPR-DSF. 25 $\mu$l of sample at 0.5 mg/ml was pipetted into 384-well plates. Temperature ramps were performed betweeen 20 \textdegree C and 100 \textdegree C at a rate of 1 \textdegree C per minute. A high power 280 nm LED was used for excitation, and the resulting fluorescence spectra were measured for wavelengths between 310 nm and 420 nm with a 25 ms integration time. The ratio of fluorescence intensities at 355 and 330 nm was calculated at each temperature. 
\section{Acknowledgements}
This work was made possible by funding from EPSRC (EP/T517872/1) and Immunocore. We would like to thank Applied Photophysics for use of the SuprDSF and continued instrument support. The Authors thank Thomas Jowitt for assistance with AUC measurements.

\bibliography{references}

\end{document}